\documentclass[11pt,a4paper]{article}
\usepackage{lmodern}
\usepackage[T1]{fontenc}
\usepackage[latin9]{inputenc}
\usepackage{amsmath}
\usepackage{amssymb}
\usepackage{stmaryrd}
\usepackage{esint}

\makeatletter

\providecommand{\tabularnewline}{\\}

\pdfoutput=1 

\usepackage{jheppub}

\usepackage{etoolbox}
    \makeatletter
    \patchcmd{\maketitle}{\@fpheader}{}{}{}
    \makeatother

\usepackage{color}

\providecommand{\tabularnewline}{\\}

\usepackage{amsfonts}

\usepackage{bbm}

\title{\boldmath Asymptotic symmetries in Carrollian theories of gravity}

\author{Alfredo P\'{e}rez}

\affiliation{Centro de Estudios Cient\'{i}ficos (CECs), Avenida Arturo Prat 514, Valdivia,
Chile.}

\emailAdd{aperez@cecs.cl}

\preprint{CECS-PHY-21/03}%

\abstract{Asymptotic symmetries in Carrollian gravitational theories in 3+1 space
 and time dimensions obtained from ``magnetic'' and ``electric'' ultrarelativistic
 contractions of General Relativity are analyzed. In both cases, parity conditions
 are needed to guarantee a finite symplectic term, in analogy with
 Einstein gravity. For the magnetic contraction, when Regge-Teitelboim parity 
conditions are imposed, the asymptotic symmetries are described by the
 Carroll group. With Henneaux-Troessaert parity conditions,
 the asymptotic symmetry algebra corresponds to a BMS-like extension of the
 Carroll algebra. For the electric contraction, because the lapse
 function does not appear in the boundary term needed to ensure a well-defined
 action principle, the
 asymptotic symmetry algebra is truncated, for Regge-Teitelboim parity conditions, to the semidirect sum of spatial
 rotations and spatial translations. Similarly, with Henneaux-Troessaert parity
 conditions, the asymptotic symmetries are given by the semidirect
 sum of spatial rotations and an infinite number of parity odd 
supertranslations. Thus, from the point of view of the asymptotic symmetries,
 the magnetic contraction can be seen as a smooth limit of General Relativity, 
in contrast to its electric counterpart.
}

\makeatother

\begin{document}
\maketitle


\section{Introduction\label{sec:1 Introduction}}

The Carroll symmetry was introduced by Levy-Leblond in 1965 as an
``ultrarelativistic'' limit of the Poincar\'e symmetry, i.e., the
limit in which the speed of light vanishes ($c\rightarrow0$) \cite{levy1965nouvelle}.
The Carroll algebra belongs to the family of Lie algebras associated
with the possible \textquotedblleft kinematical groups\textquotedblright{}
classified in ref. \cite{Bacry:1968zf}, and has been used in very
different physical contexts (see e.g. \cite{dauutcourt1967characteristic, Fareghbal:2013ifa,Dautcourt:1997hb,Duval:2014uoa,Duval:2014uva,Bergshoeff:2014jla,Fareghbal:2014qga,Cardona:2016ytk,Bergshoeff:2016soe,Duval:2017els,Grumiller:2017sjh,Ciambelli:2018wre,Figueroa-OFarrill:2018ilb,Barducci:2018thr,Morand:2018tke,Penna:2018gfx,Bagchi:2019xfx,Donnay:2019jiz,Figueroa-OFarrill:2019sex,Ravera:2019ize,Ciambelli:2019lap,Gomis:2019nih,Bagchi:2019clu,Banerjee:2020qjj,Grumiller:2020elf,Gomis:2020wxp,Bagchi:2021qfe,Concha:2021jnn,Casalbuoni:2021fel,Pena-Benitez:2021ipo,Azarnia:2021uch,Campoleoni:2021blr,Marsot:2021tvq}
and references therein). In this ultrarelativistic limit, light cones
close up, and as a consequence neighboring points become causally
disconnected. In the words of Levy-Leblond:

``\emph{In a world whose invariance group would be this new group,
there would be practically no causality}.''

It has recently been shown that relativistic field theories generically
admit two inequivalent contractions that lead to Carroll invariant
field theories \cite{Henneaux:2021yzg}. They were termed ``electric''
and ``magnetic'' contractions, respectively. They were known in
the case of electrodynamics \cite{Duval:2014uoa}, but remarkably
they also exist in theories that are not necessarily invariant under
duality transformations. Both contractions, in the case of scalar
and electromagnetic fields, were also discussed in ref. \cite{deBoer:2021jej}.

In the context of gravitation, the theory obtained from the electric
contraction of General Relativity has been known for a long time,
and can be interpreted as a strong coupling limit of Einstein gravity
\cite{Isham:1975ur}, or as a gravitational theory in a ``zero-signature
limit'' \cite{Teitelboim:1978wv}. In ref. \cite{Henneaux:1979vn}
it was written in terms of a covariant Lagrangian under Carroll transformations,
and in refs. \cite{Teitelboim:1981ua,Henneaux:1981su,Teitelboim:1983fi}
it was used as a starting point of an alternative perturbative approach
in terms of the signature parameter for the quantization of the gravitational
field, mimicking the quantization process of a relativistic free particle.
By virtue of their ``ultra-local'' properties, this theory can also
be used in the description of the spacetime near space-like singularities
\cite{Belinsky:1970ew,Belinsky:1982pk,henneaux1982quantification,Damour:2002et}.
On the other hand, the magnetic contraction was recently introduced
by Henneaux and Salgado-Rebolledo in ref. \cite{Henneaux:2021yzg},
and is described in Hamiltonian form\footnote{Different actions for gravitational theories with Carrollian symmetries
were constructed in refs. \cite{Hartong:2015xda,Bergshoeff:2017btm}
by gauging the Carroll algebra. The relation with the electric and
magnetic contractions discussed in ref. \cite{Henneaux:2021yzg} is
still unclear. However, in the same reference it was conjectured that
the electric contraction could be related with the action in \cite{Hartong:2015xda},
while the magnetic contraction to the one introduced in \cite{Bergshoeff:2017btm}.}.
\begin{table*}
\protect\caption{\label{tab:table1}General Relativity and Electric/Magnetic Carrollian
gravity compared and contrasted}

\scriptsize{

\begin{tabular}{cccccc}
 &  &  & \multicolumn{3}{l}{}\tabularnewline
\hline 
\hline 
 &  &  & \multicolumn{3}{c}{}\tabularnewline
 &  &  & \multicolumn{3}{c}{}\tabularnewline
 & General Relativity & \multicolumn{1}{c}{Magnetic Carrollian gravity} & \multicolumn{3}{c}{Electric Carrollian gravity}\tabularnewline
\hline 
 &  &  & \multicolumn{3}{c}{}\tabularnewline
 &  &  & \multicolumn{3}{c}{}\tabularnewline
Action principle & $I=\int dtd^{3}x\,(\pi^{ij}\dot{g}_{ij}-N\mathcal{H}$ & $I=\int dtd^{3}x\,(\pi^{ij}\dot{g}_{ij}-N\mathcal{H}^{M}$ & \multicolumn{3}{c}{$I=\int dtd^{3}x\,(\pi^{ij}\dot{g}_{ij}-N\mathcal{H}^{E}$}\tabularnewline
in Hamiltonian & $\quad-N^{i}\mathcal{H}_{i}$) & \multicolumn{1}{c}{$\quad-N^{i}\mathcal{H}_{i}^{M})$} & \multicolumn{3}{c}{$\quad-N^{i}\mathcal{H}_{i}^{E}$)}\tabularnewline
form &  &  & \multicolumn{3}{c}{}\tabularnewline
 & $\mathcal{H}=\frac{1}{\sqrt{g}}\left(\pi^{ij}\pi_{ij}-\frac{1}{2}\pi^{2}\right)-\sqrt{g}R$ & $\mathcal{H}^{M}=-\sqrt{g}R$ & \multicolumn{3}{c}{$\mathcal{H}^{E}=\frac{1}{\sqrt{g}}\left(\pi^{ij}\pi_{ij}-\frac{1}{2}\pi^{2}\right)$}\tabularnewline
 &  &  & \multicolumn{3}{c}{}\tabularnewline
 & $\mathcal{H}_{i}=-2\pi_{i\mid j}^{\;j}$ & $\mathcal{H}_{i}^{M}=-2\pi_{i\mid j}^{\;j}$ & \multicolumn{3}{c}{$\mathcal{H}_{i}^{E}=-2\pi_{i\mid j}^{\;j}$}\tabularnewline
 &  &  & \multicolumn{3}{c}{}\tabularnewline
 &  &  & \multicolumn{3}{c}{}\tabularnewline
Poisson bracket & $\left\{ \mathcal{H}(x),\mathcal{H}(x')\right\} =(g^{ij}\left(x\right)\mathcal{H}_{j}\left(x\right)$ & $\left\{ \mathcal{H}^{M}(x),\mathcal{H}^{M}(x')\right\} =0$ & \multicolumn{3}{c}{$\left\{ \mathcal{H}^{E}(x),\mathcal{H}^{E}(x')\right\} =0$}\tabularnewline
between two & $\quad\quad+g^{ij}\left(x'\right)\mathcal{H}_{j}\left(x'\right))\delta,_{i}\left(x,x'\right)$ &  & \multicolumn{3}{c}{}\tabularnewline
Hamiltonian &  &  & \multicolumn{3}{c}{}\tabularnewline
constraints &  &  & \multicolumn{3}{c}{}\tabularnewline
 &  &  & \multicolumn{3}{c}{}\tabularnewline
 &  &  &  &  & \tabularnewline
Boundary term & $\delta H=\oint d^{2}s_{l}\left[2N_{k}\delta\pi^{kl}\right.$ & $\delta H_{M}=\oint d^{2}s_{l}\left[2N_{k}\delta\pi^{kl}\right.$ & \multicolumn{3}{c}{$\delta H_{E}=\oint d^{2}s_{l}\left[2N_{k}\delta\pi^{kl}\right.$}\tabularnewline
in the variation of & $+(2N^{k}\pi^{jl}-N^{l}\pi^{jk})\delta g_{jk}$ & $+(2N^{k}\pi^{jl}-N^{l}\pi^{jk})\delta g_{jk}$ & \multicolumn{3}{c}{$\left.+(2N^{k}\pi^{jl}-N^{l}\pi^{jk})\delta g_{jk}\right]$}\tabularnewline
the Hamiltonian & $\left.+G^{ijkl}(N\delta g_{ij\mid k}-N_{\mid k}\delta g_{ij})\right]$ & $\left.+G^{ijkl}(N\delta g_{ij\mid k}-N_{\mid k}\delta g_{ij})\right]$ & \multicolumn{3}{c}{There is no $N$ (lapse)}\tabularnewline
 &  &  & \multicolumn{3}{c}{in the expression}\tabularnewline
 &  &  & \multicolumn{3}{c}{}\tabularnewline
 &  &  & \multicolumn{3}{c}{}\tabularnewline
Need for parity & Yes & Yes & \multicolumn{3}{c}{Yes}\tabularnewline
conditions? &  &  & \multicolumn{3}{c}{}\tabularnewline
 &  &  & \multicolumn{3}{c}{}\tabularnewline
 &  &  & \multicolumn{3}{c}{}\tabularnewline
Asymptotic & Poincar\'e algebra & Carroll algebra & \multicolumn{3}{c}{Spatial rotations$\;\subsetplus$}\tabularnewline
symmetry algebra &  &  & \multicolumn{3}{c}{spatial translations}\tabularnewline
with Regge- &  &  & \multicolumn{3}{c}{}\tabularnewline
Teitelboim &  &  & \multicolumn{3}{c}{}\tabularnewline
parity conditions &  &  & \multicolumn{3}{c}{}\tabularnewline
 &  &  & \multicolumn{3}{c}{}\tabularnewline
 &  &  & \multicolumn{3}{c}{}\tabularnewline
 &  &  & \multicolumn{3}{c}{}\tabularnewline
Asymptotic & BMS algebra & BMS-like extension & \multicolumn{3}{c}{Spatial rotations$\;\subsetplus$}\tabularnewline
symmetry algebra &  & of the Carroll algebra & \multicolumn{3}{c}{parity odd supertranslations}\tabularnewline
with Henneaux & (in an unconventional &  & \multicolumn{3}{c}{}\tabularnewline
-Troessaert parity & basis) &  & \multicolumn{3}{c}{(spatial translations}\tabularnewline
conditions &  &  & \multicolumn{3}{c}{obtained from the}\tabularnewline
 &  &  & \multicolumn{3}{c}{modes with $\ell=1$}\tabularnewline
 &  &  & \multicolumn{3}{c}{in the spherical harmonics}\tabularnewline
 &  &  & \multicolumn{3}{c}{expansion)}\tabularnewline
\hline 
\end{tabular}

}
\end{table*}

Both contractions share a common characteristic: the algebra of first
class constraints possesses a Carroll structure. In other words, the
normal surface deformations form an abelian subgroup, in sharp contrast
with the surface deformation algebra of General Relativity \cite{Teitelboim:1972vw,Teitelboim:1973yj}.
However, the fact that the algebra of constraints has a Carroll structure
in the sense described above, is not enough to guarantee that the
theory is in fact Carrollian. The theory must have a Carroll symmetry
(or an extension of it) emerging as an asymptotic symmetry defined
by appropriate improper (large) gauge transformations that change
the physical state \cite{Benguria:1976in}. Therefore, it is of fundamental
importance to study the asymptotic structure of these theories, and
to determine which of them possess Carrollian asymptotic symmetries
with well-defined canonical generators. This is the main purpose of
this article.

As in the case of General Relativity, it will be necessary to introduce
suitable parity conditions to have a finite symplectic term. Two possibilities
are explored, analogous to Regge-Teitelboim \cite{Regge:1974zd} and
Henneaux-Troessaert \cite{Henneaux:2018cst} parity conditions at
spatial infinity. Thus, the asymptotic symmetry algebra will depend
on the type of contraction, electric or magnetic, as well as on the
choice of parity conditions. In particular, for the magnetic contraction,
the asymptotic symmetries are described by the finite-dimensional
Carroll group when Regge-Teitelboim parity conditions are imposed.
With Henneaux-Troessaert parity conditions, the asymptotic symmetry
algebra corresponds to a BMS-like extension of the Carroll algebra.
On the other hand, in the case of the electric contraction, and because
the lapse function does not appear in the boundary term needed to
ensure a well-defined action principle, the asymptotic symmetry algebra
is truncated to the semi-direct sum of spatial rotations and spatial
translations when Regge-Teitelboim parity conditions are used. Similarly,
with Henneaux-Troessaert parity conditions, the asymptotic symmetries
are given by the semi-direct sum of spatial rotations and an infinite
number of parity odd supertranslations. Hence, the Carroll group is
not present in the electric contraction, and consequently, there is
no generator associated with time translations that allow defining
energy in this theory. In this sense, from the point of view of the
asymptotic symmetries, the magnetic contraction can be seen as a smooth
limit of General Relativity, in contrast to its electric counterpart.
The results are summarized and compared with the standard ones in
General Relativity in Table \ref{tab:table1}.

The plan of the paper is the following: In the next section, the asymptotic
symmetries of the theory obtained from the magnetic Carrollian contraction
of General Relativity are studied. In particular, we consider the
cases when Regge-Teitelboim parity conditions and Henneaux-Troessaert
parity conditions are implemented. The asymptotic symmetry algebra
is then determined, and the canonical realization of the charges is
obtained for each case. As an example, the charges associated with
a ``Carrollian Schwarzschild-like'' configuration are computed explicitly.
In section \ref{sec:Asymptotic-symmetries-in_Electric}, a similar
analysis is performed for the case of the ``Electric Carrollian gravity.''
Both, Regge-Teitelboim and Henneaux-Troessaert parity conditions are
considered, and the asymptotic symmetry algebra together with their
canonical generators are determined for each case. The charges associated
with a ``Carrollian Schwarzschild-like'' solution of the theory
are also computed. Section \ref{sec:5 Final-remarks} is devoted to
conclusions and final remarks. Finally, in appendix \ref{sec:Appendix A},
the results obtained in section \ref{sec:Asymptotic-symmetries-in_Magnetic}
and section \ref{sec:Asymptotic-symmetries-in_Electric} for the case
of Regge-Teitelboim parity conditions are reviewed in Cartesian coordinates.

\section{Asymptotic symmetries in Magnetic Carrollian gravity\label{sec:Asymptotic-symmetries-in_Magnetic}}

\subsection{Action principle, variation of the charge and transformation laws}

The Magnetic Carrollian theory of gravity was recently introduced
in ref. \cite{Henneaux:2021yzg}. Its action principle is obtained
from the action of General Relativity in Hamiltonian form by the so-called
``magnetic contraction,'' and is described in the canonical formalism.
The canonical action is given by
\begin{equation}
I=\int dtd^{3}x\left(\pi^{ij}\dot{g}_{ij}-N\mathcal{H}^{M}-N^{i}\mathcal{H}_{i}^{M}\right)\,,\label{eq:action}
\end{equation}
where
\begin{align}
\mathcal{H}^{M} & =-\sqrt{g}R\,,\qquad\qquad\mathcal{H}_{i}^{M}=-2\pi_{i\mid j}^{\;j}\,.\label{eq:Constraints}
\end{align}
The canonical variables are given by $g_{ij}$ and $\pi^{ij}$, with
$i,j=1,2,3$. Here $g_{ij}$ are the spatial metric components of
the corresponding four-dimensional Carrollian metric \cite{Henneaux:1979vn},
and $\pi^{ij}$ denotes their conjugate momenta. In analogy with General
Relativity, the functions $N$ and $N^{i}$ will be called ``lapse''
and ``shift'' functions, respectively. These are the Lagrange multipliers
that implement the constraints $\mathcal{H}^{M}\approx0$ and $\mathcal{H}_{i}^{M}\approx0$
defined in eq. \eqref{eq:Constraints}, where $R$ is the Ricci scalar
of the three-dimensional metric $g_{ij}$, and $\mid$ denotes covariant
differentiation with respect to this metric.

The first class constraints obey the following surface deformation
algebra
\begin{align}
\left\{ \mathcal{H}^{M}\left(x\right),\mathcal{H}^{M}\left(x'\right)\right\}  & =0\,,\label{eq:HpHp}\\
\left\{ \mathcal{H}^{M}\left(x\right),\mathcal{H}_{i}^{M}\left(x'\right)\right\}  & =\mathcal{H}^{M}\left(x\right)\delta_{,i}\left(x,x'\right)\,,\label{eq:HpHi}\\
\left\{ \mathcal{H}_{i}^{M}\left(x\right),\mathcal{H}_{j}^{M}\left(x'\right)\right\}  & =\mathcal{H}_{i}^{M}\left(x'\right)\delta_{,j}\left(x,x'\right)+\mathcal{H}_{j}^{M}\left(x\right)\delta_{,i}\left(x,x'\right)\,.\label{eq:HiHj}
\end{align}
Note that the Poisson bracket between two $\mathcal{H}^{M}$ in eq.
\eqref{eq:HpHp} vanishes, while the Poisson brackets in eqs. \eqref{eq:HpHi}
and \eqref{eq:HiHj} coincide with those of General Relativity \cite{Teitelboim:1972vw}.
Indeed, the above algebra can be obtained from the ``zero signature
limit'' of the surface deformation algebra of Einstein gravity when
the parameter defining the signature is set to zero, i.e., when $\epsilon=0$
in eqs. (28.a)-(28.c) of ref. \cite{Teitelboim:1972vw}. Furthermore,
as it was pointed out in \cite{Teitelboim:1978wv}, the commutations
relations \eqref{eq:HpHp}-\eqref{eq:HiHj} define a true algebra
in the sense that the structure constants are independent of the fields,
in contrast to General Relativity (see Table \ref{tab:table1}). The
abelian subalgebra spanned by $\mathcal{H}^{M}\left(x\right)$ is
characteristic of the Carroll symmetry because the generators that
are associated with normal deformations, i.e., the generators of time
translations and boosts, commute among them.

Hamilton equations follow directly from the action principle \eqref{eq:action}.
They read

\begin{equation}
\dot{g}_{ij}=N_{i\mid j}+N_{j\mid i}\,,\label{eq:Hamg}
\end{equation}
\begin{equation}
\dot{\pi}^{ij}=-N\sqrt{g}\left(R^{ij}-\frac{1}{2}g^{ij}R\right)+\sqrt{g}\left(N^{\mid i\mid j}-g^{ij}N_{\mid k}^{\,\mid k}\right)+\left(N^{k}\pi^{ij}\right)_{\mid k}-N_{\mid k}^{i}\pi^{kj}-N_{\mid k}^{j}\pi^{ki}\,.\label{eq:Hampi}
\end{equation}
The momenta cannot be eliminated in terms of time derivatives of $g_{ij}$,
as it can be directly seen from eq. \eqref{eq:Hamg}. This is because
$\mathcal{H}^{M}$ is independent of the momenta, while $\mathcal{H}_{i}^{M}$
is linear in them.

The generator of gauge symmetries takes the form
\begin{equation}
G\left[\xi,\xi^{i}\right]=\int d^{3}x\left(\xi\,\mathcal{H}^{M}+\xi^{i}\,\mathcal{H}_{i}^{M}\right)+Q_{M}\,,\label{eq:Generators}
\end{equation}
where, according to Regge and Teitelboim, $Q_{M}$ is the boundary
term that must be added in order to guarantee that the canonical generators
have well-defined functional derivatives \cite{Regge:1974zd}. Its
variation then reads
\begin{equation}
\delta Q_{M}=\oint d^{2}s_{l}\left[G^{ijkl}\left(\xi\delta g_{ij\mid k}-\xi_{\mid k}\delta g_{ij}\right)+2\xi_{k}\delta\pi^{kl}+\left(2\xi^{k}\pi^{jl}-\xi^{l}\pi^{jk}\right)\delta g_{jk}\right]\,,\label{eq:deltaQ}
\end{equation}
where $G^{ijkl}$ is the inverse of the de Witt supermetric, given
by 
\[
G^{ijkl}=\frac{1}{2}\sqrt{g}\left(g^{ik}g^{jl}+g^{il}g^{jk}-2g^{ij}g^{kl}\right)\,.
\]
Note that the expression for the variation of the charge \eqref{eq:deltaQ}
precisely coincides with the one in General Relativity. The reason
is that the term that was removed from $\mathcal{H}$ in the process
of contraction, $G_{ijkl}\pi^{ij}\pi^{kl}$, does not contain spatial
derivatives, and therefore it does not contribute to the boundary
term.

The transformation laws for the canonical variables $g_{ij}$ and
$\pi^{ij}$ are then generated by \eqref{eq:Generators}. They are
given by
\begin{equation}
\delta g_{ij}=\xi_{i\mid j}+\xi_{j\mid i}\,,\label{eq:transfg}
\end{equation}
\begin{equation}
\delta\pi^{ij}=-\xi\sqrt{g}\left(R^{ij}-\frac{1}{2}g^{ij}R\right)+\sqrt{g}\left(\xi^{\mid i\mid j}-g^{ij}\xi_{\mid k}^{\,\mid k}\right)+\left(\xi^{k}\pi^{ij}\right)_{\mid k}-\xi_{\mid k}^{i}\pi^{kj}-\xi_{\mid k}^{j}\pi^{ki}\,.\label{eq:transfpi}
\end{equation}
The transformation laws take exactly the same form as the Hamilton
equations in \eqref{eq:Hamg}, \eqref{eq:Hampi} with the replacement
$\xi\rightarrow N$, $\xi^{i}\rightarrow N^{i}$. This is due to the
fact that the time evolution is a gauge transformation.

\subsection{Asymptotic behavior of the fields}

For the analysis of the asymptotic symmetries it is necessary to specify
the behavior of the fields, in this case the canonical variables $g_{ij}$
and $\pi^{ij}$, in the asymptotic region. Then one has to find the
most general form of the parameters $\xi$, $\xi^{i}$ that preserves
the fall-off of the canonical variables under the set of transformations
defined in eqs. \eqref{eq:transfg} and \eqref{eq:transfpi}. The
transformations with associated non-trivial charges will then define
the asymptotic symmetries of the theory. In addition, the asymptotic
behavior of the fields must guarantee that the charges are finite
and integrable in the functional sense, and that the symplectic term
is finite.

\subsubsection{Fall-off of the canonical variables}

To describe the asymptotic behavior of the fields, we will consider
deviations with respect to the ``Carrollian background configuration''
characterized by the canonical variables $\bar{g}_{ij}$, $\bar{\pi}^{ij}$
given by
\begin{equation}
\bar{g}_{ij}dx^{i}dx^{j}=dr^{2}+r^{2}\gamma_{AB}dx^{A}dx^{B}\;,\qquad\bar{\pi}^{ij}=0\,,\label{eq:background}
\end{equation}
with $A,B=1,2$, and where $\gamma_{AB}$ denotes the metric of the
round 2-sphere. Since $\bar{g}_{ij}$ corresponds to the metric of
the three-dimensional Euclidean flat space, this configuration automatically
solves the constraints in eq. \eqref{eq:Constraints}.

In what follows, we will use spherical coordinates because they will
allow us to discuss the parity conditions of Regge-Teitelboim \cite{Regge:1974zd}
and Henneaux-Troessaert \cite{Henneaux:2018cst} using the same expressions.
The analysis in Cartesian coordinates for the case of Regge-Teitelboim
parity conditions is discussed in Appendix \ref{sec:Appendix A}.

The proposed fall-off for the canonical variables is given by

\begin{align}
g_{rr} & =1+\frac{f_{rr}}{r}+\frac{f_{rr}^{\left(-2\right)}}{r^{2}}+\mathcal{O}\left(r^{-3}\right)\,,\label{eq:grr}\\
g_{rA} & =\frac{f_{rA}^{\left(-1\right)}}{r}+\mathcal{O}\left(r^{-2}\right)\,,\label{eq:grA}\\
g_{AB} & =r^{2}\gamma_{AB}+r\,f_{AB}+f_{AB}^{\left(0\right)}+\mathcal{O}\left(r^{-1}\right)\,,\label{eq:gAB}
\end{align}
\begin{align}
\pi^{rr} & =p^{rr}+\mathcal{O}\left(r^{-1}\right)\,,\label{eq:prr}\\
\pi^{rA} & =\frac{p^{rA}}{r}+\frac{p_{\left(-2\right)}^{rA}}{r^{2}}+\mathcal{O}\left(r^{-3}\right)\,,\label{eq:prA}\\
\pi^{AB} & =\frac{p^{AB}}{r^{2}}+\mathcal{O}\left(r^{-3}\right)\,.\label{eq:pAB}
\end{align}
This is exactly the same asymptotic behavior for the fields used in
the case of General Relativity \cite{Regge:1974zd,Henneaux:2018cst}
and guarantees that the charges are finite and integrable in the functional
sense. The $\mathcal{O}\left(1\right)$ term of $g_{rA}$ has been
set zero as in ref. \cite{Henneaux:2018cst} to avoid difficulties
with the integrability. Note that all the terms that are relevant
for the charges have been explicitly displayed.

The asymptotic conditions \eqref{eq:grr}-\eqref{eq:pAB} are preserved
under the transformations given in eqs. \eqref{eq:transfg} and \eqref{eq:transfpi}
with gauge parameters of the form
\begin{align}
\xi & =r\,b+f\left(\theta,\text{\ensuremath{\phi}}\right)+\mathcal{O}\left(r^{-1}\right)\,,\label{eq:xi}\\
\xi^{r} & =W\left(\theta,\text{\ensuremath{\phi}}\right)+\mathcal{O}\left(r^{-1}\right)\,,\label{eq:xir}\\
\xi^{A} & =Y^{A}+\frac{\partial^{A}W\left(\theta,\text{\ensuremath{\phi}}\right)}{r}+\mathcal{O}\left(r^{-2}\right)\,,\label{eq:xiA}
\end{align}
with
\begin{equation}
b=\vec{\beta}\cdot\hat{r}\;,\qquad Y^{A}=\frac{\epsilon^{AB}}{\sqrt{\gamma}}\partial_{B}\left(\vec{\omega}\cdot\hat{r}\right)\,.\label{bY}
\end{equation}
Here, $\epsilon^{AB}$ is the antisymmetric Levi-Civita symbol and
$\gamma$ denotes the determinant of the 2-sphere metric $\gamma_{AB}$.
The indices $A,B$ are lowered and raised with the same metric. The
vector $\hat{r}=\left(\sin\theta\cos\phi,\sin\theta\sin\phi,\cos\theta\right)$
is the unit normal to the 2-sphere, and the three-dimensional
constant vectors $\vec{\beta}$ and $\vec{\omega}$ are the parameters
associated with boosts and spatial rotations, respectively. The functions
$f=f\left(\theta,\text{\ensuremath{\phi}}\right)$ and $W=W\left(\theta,\phi\right)$
are general arbitrary scalar functions defined on the 2-sphere.

\subsubsection{Variation of the charge}

The variation of the charge is obtained by replacing the asymptotic
conditions \eqref{eq:grr}-\eqref{eq:pAB} and the asymptotic form
of the gauge parameters \eqref{eq:xi}-\eqref{bY} in the expression
\eqref{eq:deltaQ} for $\delta Q_{M}$. It contains a divergent term
in the large $r$ expansion that can be eliminated by using the order
$\mathcal{O}\left(r^{-1}\right)$ of the constraint $\mathcal{H}^{M}$
given by
\begin{equation}
\Delta f_{rr}+\Delta\tilde{f}-D^{A}D^{B}f_{AB}=0\,,\label{eq:Const3}
\end{equation}
 and the order $\mathcal{O}\left(1\right)$ of the constraint $\mathcal{H}_{A}^{M}$
\begin{equation}
p^{rA}+D_{B}p^{AB}=0\,,\label{eq:const1}
\end{equation}
as it was done by Henneaux and Troessaert in the case of General Relativity
\cite{Henneaux:2018cst}. Here $D_{A}$ denotes the covariant derivative
with respect to the metric of the 2-sphere $\gamma_{AB}$, and $\Delta=D_{A}D^{A}$
is the corresponding Laplacian. We are using the notation $\tilde{X}:=\gamma^{AB}X_{AB}$
to represent the trace of a tensor $X_{AB}$.

The divergent part of the charge reads
\[
Q_{M}^{\text{div}}=r\oint d^{2}x\,2\sqrt{\gamma}\left[\left(\vec{\beta}\cdot\hat{r}\right)\left(f_{rr}+\frac{1}{2}\tilde{f}\right)+p^{rA}\epsilon_{AC}D^{C}\left(\vec{\omega}\cdot\hat{r}\right)\right]\,.
\]
 Using the constraints \eqref{eq:Const3}, \eqref{eq:const1} and
performing appropriate integrations by parts, one can show that the
divergent term vanishes identically
\[
Q_{M}^{\text{div}}=0\,.
\]

The finite part of the variation of the charge is given by
\begin{align}
\delta Q_{M} & =\delta\oint d^{2}x\left[f\left(2\sqrt{\gamma}f_{rr}\right)+2Y^{A}\left(p_{\;A\left(-2\right)}^{r}+f_{AB}p^{rB}\right)+2W\left(p^{rr}-D_{A}p^{rA}\right)\right.\nonumber \\
 & \left.+\sqrt{\gamma}\,b\left(2f_{rr}^{\left(-2\right)}+2D^{A}f_{rA}^{\left(-1\right)}+2\tilde{f}^{\left(0\right)}-\frac{3}{2}f_{rr}^{2}-\frac{3}{4}f^{AB}f_{AB}+\frac{\tilde{f}^{2}}{4}\right)\right]\nonumber \\
 & +\oint d^{2}x\frac{\sqrt{\gamma}\,b}{2}\left(\tilde{f}\delta f_{rr}-f_{rr}\delta\tilde{f}\right)\,.\label{eq:DeltaQM}
\end{align}
This expression contains a non-integrable part that will be eliminated
by the parity conditions.

\subsubsection{Transformation laws}

The transformation laws of the fields that are relevant for the discussion
of the parity conditions are the following:

\begin{equation}
\delta f_{rr}=\mathcal{L}_{Y}f_{rr}\,,\label{eq:deltafrr}
\end{equation}
\begin{equation}
\delta f_{AB}=\mathcal{L}_{Y}f_{AB}+2\left(D_{A}D_{B}+\gamma_{AB}\right)W\,,\label{eq:deltafAB}
\end{equation}
\begin{align}
\delta p^{rr} & =\mathcal{L}_{Y}p^{rr}+b\frac{\sqrt{\gamma}}{2}\left[6f_{rr}+D^{A}D^{B}f_{AB}-\tilde{f}-\Delta\tilde{f}\right]+\sqrt{\gamma}\left(D^{A}b\right)\left(D^{B}f_{AB}\right)\nonumber \\
 & -\frac{\sqrt{\gamma}}{2}\left(D^{A}b\right)\left(D_{A}\tilde{f}\right)-\sqrt{\gamma}\Delta f\,,\label{eq:deltaPrr}
\end{align}
\begin{equation}
\delta p^{rA}=\mathcal{L}_{Y}p^{rA}+\frac{\sqrt{\gamma}}{2}D_{B}\left(bf^{AB}\right)-\frac{\sqrt{\gamma}}{2}bD^{A}\left(\tilde{f}+2f_{rr}\right)-\sqrt{\gamma}D^{A}f\,,\label{eq:deltaPrA}
\end{equation}

\begin{align}
\delta p^{AB} & =\mathcal{L}_{Y}p^{AB}+\frac{\sqrt{\gamma}}{2}b\left(D^{A}D^{B}f_{rr}-\gamma^{AB}\Delta f_{rr}+f^{AB}\right)\nonumber \\
 & +\frac{\sqrt{\gamma}}{2}\left[\left(D^{A}b\right)\left(D_{C}f^{CB}\right)+\left(D^{B}b\right)\left(D_{C}f^{CA}\right)\right]\nonumber \\
 & -\frac{\sqrt{\gamma}}{2}\left[\left(D^{A}b\right)\left(D^{B}\tilde{f}\right)+\left(D^{B}b\right)\left(D^{A}\tilde{f}\right)\right]-\frac{\sqrt{\gamma}}{2}\left(D^{C}b\right)\left(D_{C}f^{AB}\right)\nonumber \\
 & +\frac{\sqrt{\gamma}}{2}\gamma^{AB}\left(D^{C}b\right)D_{C}\left(\tilde{f}-f_{rr}\right)+\sqrt{\gamma}\left(D^{A}D^{B}-\gamma^{AB}\Delta\right)f\,.\label{eq:DeltaPAB}
\end{align}
Here, $\mathcal{L}_{Y}$ denotes the Lie derivative with respect to
the vector field on the sphere $Y^{A}$. Note that the field $f_{rr}$,
which in the variation of the charge \eqref{eq:DeltaQM} is associated
with the term of order $\mathcal{O}\left(1\right)$ in the time translations,
transforms only under spatial rotations in contrast to the case of
General Relativity. This is the imprint, at the level of the transformation
laws of the fields, of the Carrollian structure of the theory.

\subsubsection{Symplectic term\label{subsec:Symplectic-term}}

For large values of $r$, the symplectic term takes the form
\begin{equation}
\int dtd^{3}x\,\pi^{ij}\dot{g}_{ij}\underset{r\rightarrow\infty}{=}\log\left(r\right)\int dt\oint d^{2}x\left(p^{rr}\dot{f}_{rr}+p^{AB}\dot{f}_{AB}\right)+\mathcal{O}\left(r^{-1}\right)\,.\label{eq:Sympmag}
\end{equation}
It possesses a logarithmic divergence that can be removed by imposing
appropriate parity conditions on the canonical variables. Since the
symplectic term is identical to the one in General Relativity, one
can use the same Regge-Teitelboim \cite{Regge:1974zd} and Henneaux-Troessaert
\cite{Henneaux:2018cst} parity conditions that were used in Einstein
gravity.

\subsection{Asymptotic symmetries with Regge-Teitelboim parity conditions}

\subsubsection{Parity conditions}

Regge-Teitelboim parity conditions were introduced in ref. \cite{Regge:1974zd}
in the context of the study of asymptotically flat spacetimes in General
Relativity. In that case, they cancel the logarithmic divergence in
the symplectic term, and the asymptotic symmetry algebra reduces to
the Poincar\'e algebra. In this section, the same Regge-Teitelboim parity
condition will be used to remove the divergence in the symplectic
term \eqref{eq:Sympmag} of the magnetic Carrollian theory.

In spherical coordinates, the antipodal map on the 2-sphere is obtained
from the following transformation 
\begin{align*}
\theta & \rightarrow-\theta+\pi\,,\\
\phi & \rightarrow\phi+\pi\,.
\end{align*}
The Regge-Teitelboim parity condition in spherical coordinates are
then given by
\[
f_{ij}dx^{i}dx^{j}\quad\text{(even under the antipodal map)}\,,
\]

\[
p^{ij}\partial_{i}\partial_{j}\quad\text{(odd under the antipodal map)}\,.
\]
Explicitly, they read
\begin{equation}
f_{rr},\,f_{\theta\theta},\,f_{\phi\phi},\,p^{\theta\phi},\,p^{r\theta}\qquad\text{(parity even)\,,}\label{eq:parityRTM1}
\end{equation}
\begin{equation}
f_{\theta\phi},\,p^{rr},\,p^{\theta\theta},\,p^{\phi\phi},\,p^{r\phi}\qquad\text{(parity odd)\,.}\label{eq:parityRTM2}
\end{equation}
The logarithmic divergence in the symplectic term in eq. \eqref{eq:Sympmag}
only contains fields of opposite parity. Therefore, it vanishes after
the integration on the 2-sphere.

For consistency, the parity conditions \eqref{eq:parityRTM1}, \eqref{eq:parityRTM2}
must be preserved under the transformation laws of the fields, imposing
additional restrictions on the gauge parameters.

From the transformation of $f_{AB}$ in eq. \eqref{eq:deltafAB},
$W$ must be even, with the exception of the mode that satisfies $\left(D_{A}D_{B}+\gamma_{AB}\right)W=0$.
From the trace of this equation, it becomes clear that the mode with
$\ell=1$ in the expansion in spherical harmonics obeys this restriction.
Thus, one has
\[
W=\vec{\alpha}\cdot\hat{r}+W_{\text{even}}\left(\theta,\phi\right)\,.
\]
Here, $\vec{\alpha}$ is a constant vector field that will be associated
with spatial translations, and $W_{\text{even}}$ contains only spherical
harmonics of even parity, i.e., those with even values of $\ell$.

From the transformation law of $p^{rr}$ in eq. \eqref{eq:deltaPrr},
$f$ must be parity odd, with the exception of the mode that satisfies
$\Delta f=0$, i.e., the mode with $\ell=0$. Therefore,
\[
f=T+f_{\text{odd}}\left(\theta,\phi\right)\,,
\]
where $T$ is a constant, and $f_{\text{odd}}\left(\theta,\phi\right)$
contains parity odd spherical harmonics. With these restrictions on
the parameters, all the transformation laws preserve the Regge-Teitelboim
parity conditions.

\subsubsection{Charges and asymptotic symmetry algebra}

When the parity conditions \eqref{eq:parityRTM1} and \eqref{eq:parityRTM2}
are implemented in the variation of the charge \eqref{eq:DeltaQM},
all the quadratic terms vanish and the charge becomes integrable in
the functional sense. Furthermore, there are no contributions to the
charge coming from the parameters $W_{\text{even}}\left(\theta,\phi\right)$
and $f_{\text{odd}}\left(\theta,\phi\right)$, and therefore, they
define transformations that are pure gauge (proper gauge transformations).
The expression for the charge then simplifies to
\begin{align}
Q_{M} & =\oint d^{2}x\;2\sqrt{\gamma}\left[Tf_{rr}+\left(\vec{\omega}\cdot\hat{r}\right)\epsilon_{AB}D^{A}p_{\left(-2\right)}^{rB}+\frac{\left(\vec{\alpha}\cdot\hat{r}\right)}{\sqrt{\gamma}}\left(p^{rr}-D_{A}p^{rA}\right)\right.\nonumber \\
 & \left.+\left(\vec{\beta}\cdot\hat{r}\right)\left(f_{rr}^{\left(-2\right)}+D^{A}f_{rA}^{\left(-1\right)}+\tilde{f}^{\left(0\right)}\right)\right]\,.\label{eq:ChargeRT}
\end{align}
The asymptotic form of the gauge parameters expressed in terms of
the parameters associated to improper (large) gauge transformations
is then given by
\begin{align}
\xi & =r\,\left(\vec{\beta}\cdot\hat{r}\right)+T+\mathcal{O}\left(r^{-1}\right)\,,\label{eq:xi-2}\\
\xi^{r} & =\vec{\alpha}\cdot\hat{r}+\mathcal{O}\left(r^{-1}\right)\,,\label{eq:xir-2}\\
\xi^{A} & =\frac{\epsilon^{AB}}{\sqrt{\gamma}}\partial_{B}\left(\vec{\omega}\cdot\hat{r}\right)+\frac{\partial^{A}\left(\vec{\alpha}\cdot\hat{r}\right)}{r}+\mathcal{O}\left(r^{-2}\right)\,.\label{eq:xiA-2}
\end{align}
The asymptotic symmetry algebra can then be obtained using the ``Carrollian
surface deformation algebra'' \eqref{eq:HpHp}-\eqref{eq:HiHj}.
The commutator of two deformations with parameters $\left(\xi_{1},\xi_{1}^{i}\right)$
and $\left(\xi_{2},\xi_{2}^{i}\right)$ gives new parameters $\left(\xi_{3},\xi_{3}^{i}\right)$
that are given by 
\begin{align}
\xi_{3} & =\xi_{1}^{i}\partial_{i}\xi_{2}-\xi_{2}^{i}\partial_{i}\xi_{1}\,,\label{eq:complaw1}\\
\xi_{3}^{i} & =\xi_{1}^{j}\partial_{j}\xi_{2}^{i}-\xi_{2}^{j}\partial_{j}\xi_{1}^{i}\,.\label{eq:complaw2}
\end{align}
These expressions can be directly obtained by setting $\epsilon=0$
(zero signature limit) in eqs. (25.a) and (25.b) in ref. \cite{Teitelboim:1972vw}.

From the asymptotic form of the parameters \eqref{eq:xi}-\eqref{bY},
together with the expressions $W=\vec{\alpha}\cdot\hat{r}$ and $f=T$
coming from the parity conditions, we find

\begin{align*}
T_{3} & =\delta_{IJ}\left(\alpha_{1}^{I}\beta_{2}^{J}-\alpha_{2}^{I}\beta_{1}^{J}\right)\;,\qquad\alpha_{3}^{K}=-\epsilon_{IJK}\left(\alpha_{1}^{I}\omega_{2}^{J}-\alpha_{2}^{I}\omega_{1}^{J}\right)\,,\\
\beta_{3}^{K} & =-\epsilon_{IJK}\left(\beta_{1}^{I}\omega_{2}^{J}-\beta_{2}^{I}\omega_{1}^{J}\right)\;,\qquad\omega_{3}^{K}=-\epsilon_{IJK}\omega_{1}^{I}\omega_{2}^{J}\,,
\end{align*}
where the indices $I,J,K=1,2,3$ label the components of the three-dimensional
vectors $\vec{\alpha}$, $\vec{\beta}$ and $\vec{\omega}$. The above
composition rule is precisely the one of the Carroll algebra \cite{levy1965nouvelle,Bacry:1968zf}.
Indeed, if
\begin{align*}
E & =2\oint d^{2}x\;\sqrt{\gamma}f_{rr}\;,\qquad P_{I}=2\oint d^{2}x\;\hat{r}_{I}\left(p^{rr}-D_{A}p^{rA}\right)\,,\\
K_{I} & =2\oint d^{2}x\;\sqrt{\gamma}\hat{r}_{I}\left(f_{rr}^{\left(-2\right)}+D^{A}f_{rA}^{\left(-1\right)}+\tilde{f}^{\left(0\right)}\right)\;,\qquad J_{I}=2\sqrt{\gamma}\oint d^{2}x\;\hat{r}_{I}\epsilon_{AB}D^{A}p_{\left(-2\right)}^{rB}\,,
\end{align*}
the charge \eqref{eq:ChargeRT} takes the form
\[
Q_{M}=T\,E+\vec{\omega}\cdot\vec{J}+\vec{\alpha}\cdot\vec{P}+\vec{\beta}\cdot\vec{K}\,,
\]
and the generators $E$, $P_{I}$, $K_{I}$ and $J_{I}$ fulfill the
Carroll algebra with the following non-vanishing Poisson brackets
\[
\left\{ P_{I},K_{J}\right\} =\delta_{IJ}E\,,\qquad\left\{ J_{I},J_{J}\right\} =-\epsilon_{IJK}J_{K}\,,
\]
\[
\left\{ P_{I},J_{J}\right\} =-\epsilon_{IJK}P_{K}\,,\qquad\left\{ K_{I},J_{I}\right\} =-\epsilon_{IJK}K_{K}\,.
\]
Thus, the asymptotic symmetry algebra of the magnetic Carrollian theory
of gravity with Regge-Teitelboim parity conditions is precisely the
four-dimensional Carroll algebra of Levy-Leblond.

\subsection{Asymptotic symmetries with Henneaux-Troessaert parity conditions\label{subsec:Asymptotic-symmetries-with}}

\subsubsection{Parity conditions}

In the case of General Relativity, a different set of parity conditions
was recently introduced by Henneaux and Troessaert \cite{Henneaux:2018cst}
that extends the asymptotic symmetry algebra at spatial infinity from
the Poincar\'e algebra to the BMS$_{4}$ algebra \cite{Bondi:1962,Sachs:1962}.
As it was discussed in section \ref{subsec:Symplectic-term}, since
the symplectic term of the magnetic Carrollian theory takes exactly
the same form as the one in Einstein gravity, the Henneaux-Troessaert
parity conditions can also be implemented in the case of the Magnetic
Carrollian gravity. The asymptotic symmetry algebra will then be enlarged
from the Carroll algebra to a BMS-like extension of it, that includes
additional ``Carrollian supertranslations.''

Following ref. \cite{Henneaux:2018cst}, in order to implement the
Henneaux-Troessaert parity conditions it is useful to introduce the
following new variables
\[
\bar{\lambda}=\frac{1}{2}f_{rr}\;,\qquad\qquad\bar{k}_{AB}=\frac{1}{2}f_{AB}+\frac{1}{2}f_{rr}\gamma_{AB}\;,\qquad\qquad\bar{k}=\frac{1}{2}\tilde{f}+f_{rr}\,,
\]
\begin{equation}
\bar{p}=2\left(p^{rr}-p^{AB}\gamma_{AB}\right)\;,\qquad k^{\left(2\right)}=-\frac{f_{rr}\tilde{f}}{4}-\frac{3}{4}f_{rr}^{2}+f_{rr}^{\left(-2\right)}+D^{A}f_{rA}^{\left(-1\right)}+\tilde{f}_{\left(0\right)}-\frac{1}{2}f^{AB}f_{AB}\,.\label{eq:newvar}
\end{equation}
The Henneaux-Troessaert parity conditions are then defined by
\begin{equation}
\bar{\lambda},\,p^{r\phi},\,p^{\theta\theta},\,p^{\phi\phi},\,\bar{k}_{\theta\phi}\qquad\text{(parity even)\,,}\label{eq:parityHTM1}
\end{equation}
\begin{equation}
\bar{p},\,p^{r\theta},\,p^{\theta\phi},\,\bar{k}_{\theta\theta},\,\bar{k}_{\phi\phi}\qquad\text{(parity odd)\,.}\label{eq:parityHTM2}
\end{equation}
In terms of these new variables the symplectic term \eqref{eq:Sympmag}
takes the form
\begin{equation}
\int dtd^{3}x\,\pi^{ij}\dot{g}_{ij}\underset{r\rightarrow\infty}{=}\log\left(r\right)\int dt\oint d^{2}x\left(\bar{p}\dot{\bar{\lambda}}+2p^{AB}\dot{\bar{k}}_{AB}\right)+\mathcal{O}\left(r^{-1}\right)\,,\label{eq:Sympmag-1}
\end{equation}
expression that vanishes by virtue of the parity conditions \eqref{eq:parityHTM1},
\eqref{eq:parityHTM2}. Therefore, the symplectic term is finite once
the Henneaux-Troessaert parity conditions are imposed.

The parameters take the same form as in eqs. \eqref{eq:xi}-\eqref{bY},
with
\begin{equation}
f=-\frac{1}{2}b\left(3f_{rr}+\tilde{f}\right)+T\left(\theta,\phi\right)\,,\label{eq:newf}
\end{equation}
where the functions $T\left(\theta,\phi\right)$ and $W\left(\theta,\phi\right)$
have the following parity under the antipodal map 
\begin{align*}
T\left(\theta,\phi\right) & \qquad\left(\text{parity even}\right)\,,\\
W\left(\theta,\phi\right) & \qquad\left(\text{parity odd}\right)\,.
\end{align*}
The parity conditions \eqref{eq:parityHTM1}, \eqref{eq:parityHTM2}
are then preserved by the transformation laws of the fields, which
in terms of the variables introduced in eqs. \eqref{eq:newvar} read

\[
\delta\bar{\lambda}=\mathcal{L}_{Y}\bar{\lambda}\,,
\]
\[
\delta\bar{k}_{AB}=\mathcal{L}_{Y}\bar{k}_{AB}+\left(D_{A}D_{B}+\gamma_{AB}\right)W\,,
\]
\begin{align*}
\delta\bar{p} & =4\sqrt{\gamma}b\left(\Delta\bar{\lambda}+3\bar{\lambda}\right)+4\sqrt{\gamma}\left(D^{C}b\right)\left(D_{C}\bar{\lambda}\right)+\mathcal{L}_{Y}\bar{p}\,,
\end{align*}
\begin{align*}
\delta p^{AB} & =\mathcal{L}_{Y}p^{AB}+\sqrt{\gamma}\left(D^{A}D^{B}-\gamma^{AB}\Delta\right)T+\sqrt{\gamma}\,b\left(\bar{k}^{AB}-D^{A}D^{B}\bar{k}+\gamma^{AB}\left(\Delta\bar{k}-2\bar{k}\right)\right)\\
 & +\sqrt{\gamma}\left[\left(D^{A}b\right)D_{C}\bar{k}^{CB}+\left(D^{B}b\right)D_{C}\bar{k}^{CA}\right]-2\sqrt{\gamma}\left[\left(D^{A}b\right)\left(D^{B}\bar{k}\right)+\left(D^{B}b\right)\left(D^{A}\bar{k}\right)\right]\\
 & -\sqrt{\gamma}\left(D^{C}b\right)\left(D_{C}\bar{k}^{AB}\right)+3\sqrt{\gamma}\gamma^{AB}\left(D_{C}b\right)\left(D^{C}\bar{k}\right)\,,
\end{align*}
\begin{align*}
\delta p^{rA} & =\mathcal{L}_{Y}p^{rA}+\sqrt{\gamma}D_{B}\left(b\,\bar{k}^{AB}\right)+\sqrt{\gamma}\left(D^{A}b\right)\bar{k}-\sqrt{\gamma}D^{A}T\,.
\end{align*}

\subsubsection{Charges and asymptotic symmetry algebra}

With Henneaux-Troessart parity conditions the charge \eqref{eq:DeltaQM}
becomes integrable in the functional sense and simplifies to

\begin{align}
Q_{M} & =\oint d^{2}x\;\left[4T\sqrt{\gamma}\bar{\lambda}+W\,\bar{p}+2Y_{A}\,\left(p_{\left(-2\right)}^{rA}-2\bar{\lambda}p^{rA}\right)+2\sqrt{\gamma}\,b\left(k^{\left(2\right)}-3\bar{\lambda}\bar{k}\right)\right]\,.\label{eq:QMHT}
\end{align}
In contrast to the case of the Regge-Teitelboim parity conditions,
the gauge parameters in \eqref{eq:xi}-\eqref{bY} now depend explicitly
on the fields through eq. \eqref{eq:newf}. Consequently, this field
dependence must be taken into account in the computation of the asymptotic
symmetry algebra. By that reason, the commutator between two surface
deformations now contains additional terms that consider the variation
of the fields. Thus, 
\begin{align*}
\xi_{3}^{\perp} & =\xi_{1}^{i}\partial_{i}\xi_{2}^{\perp}-\xi_{2}^{i}\partial_{i}\xi_{1}^{\perp}+\delta_{2}\xi_{1}^{\perp}-\delta_{1}\xi_{2}^{\perp}\,,\\
\xi_{3}^{i} & =\xi_{1}^{j}\partial_{j}\xi_{2}^{i}-\xi_{2}^{j}\partial_{j}\xi_{1}^{i}+\delta_{2}\xi_{1}^{i}-\delta_{1}\xi_{2}^{i}\,.
\end{align*}
The algebra of the asymptotic transformations is then given by

\[
Y_{3}^{A}=Y_{1}^{B}\partial_{B}Y_{2}^{A}-\left(1\leftrightarrow2\right)\,,
\]
\[
b_{3}=Y_{1}^{A}D_{A}b_{2}-\left(1\leftrightarrow2\right)\,,
\]
\[
T_{3}=Y_{1}^{A}D_{A}T_{2}-3b_{1}W_{2}-\left(D_{A}b_{1}\right)\left(D^{A}W_{2}\right)-b_{1}\Delta W_{2}-\left(1\leftrightarrow2\right)\,,
\]

\[
W_{3}=Y_{1}^{A}D_{A}W_{2}-\left(1\leftrightarrow2\right)\,.
\]
It defines an extension of the Carroll algebra with additional symmetry
transformations characterized by the even modes with $\ell\geq2$
of $T\left(\theta,\phi\right)$, and by the odd modes with $\ell\geq3$
of $W\left(\theta,\phi\right)$ in the spherical harmonics expansion.
The modes with $\ell=0$ of $T$ and $\ell=1$ of $W$ parametrize
the symmetry transformations associated with the energy and the linear
momentum, respectively.

The canonical generators associated with the above transformations
are then given by
\begin{equation}
\mathcal{T}\left(\theta,\phi\right)=4\bar{\lambda}\;,\qquad\mathcal{P}\left(\theta,\phi\right)=\frac{\bar{p}}{\sqrt{\gamma}}\,,\label{eq:GenMagHT1}
\end{equation}
\begin{equation}
J_{I}=\oint d^{2}x\;2\sqrt{\gamma}\epsilon_{AB}\hat{r}_{I}\,D^{A}\left(p_{\left(-2\right)}^{rB}-2\bar{\lambda}p^{rB}\right)\;,\qquad K_{I}=\oint d^{2}x\;2\sqrt{\gamma}\,\hat{r}_{I}\left(k^{\left(2\right)}-3\bar{\lambda}\bar{k}\right)\,.\label{eq:GenMagHT2}
\end{equation}
Thus, the charge takes the form
\[
Q_{M}=\vec{\omega}\cdot\vec{J}+\vec{\beta}\cdot\vec{K}+\oint d^{2}x\;\sqrt{\gamma}\left[T\left(\theta,\phi\right)\,\mathcal{T}\left(\theta,\phi\right)+W\left(\theta,\phi\right)\,\mathcal{P}\left(\theta,\phi\right)\right]\,.
\]
The generators then have the following non-vanishing Poisson brackets
\[
\left\{ J_{I},J_{J}\right\} =-\epsilon_{IJK}J_{K},\qquad\left\{ K_{I},J_{I}\right\} =-\epsilon_{IJK}K_{K}\,,
\]
\[
\left\{ \mathcal{P}\left(\theta,\phi\right),J_{I}\right\} =\hat{Y}_{I}^{A}\partial_{A}\mathcal{P}\left(\theta,\phi\right)\;,\qquad\left\{ \mathcal{T}\left(\theta,\phi\right),J_{I}\right\} =\hat{Y}_{I}^{A}\partial_{A}\mathcal{\mathcal{T}}\left(\theta,\phi\right)\,,
\]
\[
\left\{ \mathcal{P}\left(\theta,\phi\right),K_{I}\right\} =\hat{r}_{I}\left(3\mathcal{T}+\Delta\mathcal{T}\right)+\left(D_{A}\hat{r}_{I}\right)\left(D^{A}\mathcal{T}\right)\,,
\]
where $\hat{Y}_{I}^{A}:=\frac{\epsilon^{AB}}{\sqrt{\gamma}}\partial_{B}\hat{r}_{I}$.
The asymptotic symmetry algebra is canonically realized and corresponds
to an extension of the Carroll algebra with additional ``Carrollian
supertranslations.''

\subsection{Example: Magnetic Carrollian Schwarzschild-like solution}

Let us consider the solution of the Carrollian magnetic theory whose
spatial metric coincides with the one of the Schwarzschild solution
in General Relativity\footnote{The Newton constant was chosen to be $G=\frac{1}{16\pi}$.}
\[
g_{rr}=\frac{1}{\left(1-\frac{M}{8\pi r}\right)}\;,\qquad\qquad g_{rA}=0\;,\qquad\qquad g_{AB}=r^{2}\gamma_{AB}\,,
\]
\[
\pi^{rr}=0\;,\qquad\qquad\pi^{rA}=0\;,\qquad\qquad\pi^{AB}=0\,.
\]
This configuration is spherically symmetric and satisfies the constraints
\eqref{eq:Constraints}. In particular, for a vanishing shift, the
dynamical equation \eqref{eq:Hampi} fixes the lapse as 
\[
N=\sqrt{1-\frac{M}{8\pi r}}\,.
\]
Note that this is exactly the same lapse as in the case of General
Relativity, but now there is no notion of a four-dimensional Riemannian
metric that can be reconstructed from the spatial metric, the lapse
and the shift.

The fields have the following asymptotic expansion:
\begin{align*}
g_{rr} & =1+\frac{M}{8\pi r}+\frac{M^{2}}{64\pi^{2}r^{2}}+\dots\,,\\
g_{AB} & =r^{2}\gamma_{AB}\,,\\
N & =1-\frac{M}{16\pi r}+\dots\,.
\end{align*}
The asymptotic behavior fits within the boundary conditions \eqref{eq:grr}-\eqref{eq:xiA}
and obeys the Regge-Teitelboim parity condition. Then, using eq. \eqref{eq:ChargeRT}
the only non-trivial charge is given by the generator
\[
E=2\oint d^{2}x\;\sqrt{\gamma}f_{rr}=M\,.
\]
Thus, the energy of this configuration with Regge-Teitelboim parity
conditions coincides with the ``Schwarzschild mass.''

In order to obtain the charges in the case of Henneaux-Troessaert
parity conditions, one must perform the following change of coordinates
$r\rightarrow r+M/\left(16\pi\right)$. With this choice one has $\bar{k}_{AB}=0$
and
\[
\bar{\lambda}=\frac{M}{16\pi}\,.
\]
The only non-trivial generator then corresponds to the zero mode of
$\mathcal{T}\left(\theta,\text{\ensuremath{\phi}}\right)=4\bar{\lambda}$
that defines the Carroll energy
\[
E=\oint d\theta d\phi\sin\theta\,\mathcal{T}=M\,.
\]

\section{Asymptotic symmetries in Electric Carrollian gravity\label{sec:Asymptotic-symmetries-in_Electric}}

\subsection{Action principle, variation of the charge and transformation laws}

The Carrollian theory obtained from the electric contraction of General
Relativity has been known for a long time. It can be understood as
the ``strong coupling limit'' of Einstein gravity ($G\rightarrow\infty$)
\cite{Isham:1975ur}, or alternatively as the limit where the signature
of the spacetime vanishes \cite{Teitelboim:1978wv}. Its action in
canonical form is given by
\begin{equation}
I=\int dtd^{3}x\left(\pi^{ij}\dot{g}_{ij}-N\mathcal{H}^{E}-N^{i}\mathcal{H}_{i}^{E}\right)\,,\label{eq:actionelectric}
\end{equation}
with
\begin{align}
\mathcal{H}^{E} & =\frac{1}{\sqrt{g}}\left(\pi^{ij}\pi_{ij}-\frac{1}{2}\pi^{2}\right),\qquad\qquad\mathcal{H}_{i}^{E}=-2\pi_{i\mid j}^{\;j}\,.\label{eq:constraints_electric}
\end{align}
The constraint $\mathcal{H}^{E}$ is quadratic in the momenta, and
does not contain derivatives of the spatial metric. In this sense,
the action \eqref{eq:actionelectric} resembles the action of a free
relativistic particle, that can be considered as a ``gravitational
theory in 1+0 dimensions.'' This was the starting point of an alternative
perturbative approach for quantum gravity \cite{Teitelboim:1981ua},
where the ``free theory'' is precisely described by the action \eqref{eq:actionelectric}.

The first class constraints obey the same Carrollian surface deformation
algebra of the magnetic contraction
\begin{align}
\left\{ \mathcal{H}^{E}\left(x\right),\mathcal{H}^{E}\left(x'\right)\right\}  & =0\,,\label{eq:HpHp-1}\\
\left\{ \mathcal{H}^{E}\left(x\right),\mathcal{H}_{i}^{E}\left(x'\right)\right\}  & =\mathcal{H}^{E}\left(x\right)\delta_{,i}\left(x,x'\right)\,,\label{eq:HpHi-1}\\
\left\{ \mathcal{H}_{i}^{E}\left(x\right),\mathcal{H}_{j}^{E}\left(x'\right)\right\}  & =\mathcal{H}_{i}^{E}\left(x'\right)\delta_{,j}\left(x,x'\right)+\mathcal{H}_{j}^{E}\left(x\right)\delta_{,i}\left(x,x'\right)\,,\label{eq:HiHj-1}
\end{align}
and the equations of motion are directly obtained from the action
\eqref{eq:actionelectric}
\begin{equation}
\dot{g}_{ij}=\frac{2N}{\sqrt{g}}\left(\pi_{ij}-\frac{1}{2}g_{ij}\pi\right)+N_{i\mid j}+N_{j\mid i}\,,\label{eq:HamEl1}
\end{equation}

\begin{eqnarray}
\dot{\pi}^{ij} & = & \frac{N}{2\sqrt{g}}g^{ij}\left(\pi^{kl}\pi_{kl}-\frac{1}{2}\pi^{2}\right)-\frac{2N}{\sqrt{g}}\left(\pi_{\;l}^{i}\pi^{lj}-\frac{1}{2}\pi\pi^{ij}\right)\nonumber \\
 &  & +\left(N^{k}\pi^{ij}\right)_{\mid k}-N_{\mid k}^{i}\pi^{kj}-N_{\mid k}^{j}\pi^{ki}\,.\label{eq:Eqpi}
\end{eqnarray}
In contrast to the magnetic contraction, using eq. \eqref{eq:HamEl1}
now it is possible to express the momenta in terms of the time derivatives
of the spatial metric.

The canonical generator associated with gauge symmetries takes the
form
\begin{equation}
G\left[\xi,\xi^{i}\right]=\int d^{3}x\left(\xi\,\mathcal{H}^{E}+\xi^{i}\,\mathcal{H}_{i}^{E}\right)+Q_{E}\,,\label{eq:Generators-1-1}
\end{equation}
where $\delta Q_{E}$ is given by
\begin{equation}
\delta Q_{E}=\oint d^{2}s_{l}\left(2\xi_{k}\delta\pi^{kl}+\left(2\xi^{k}\pi^{jl}-\xi^{l}\pi^{jk}\right)\delta g_{jk}\right)\,.\label{eq:DeltaQE}
\end{equation}
It is worth highlighting that the parameter $\xi$, that characterizes
the deformations that move outside the $t=const.$ hypersurfaces does
not appear in $\delta Q_{E}$. As it will be discussed in the next
sections, this property of the Electric Carrollian theory plays a
key role in the form of the asymptotic symmetry algebra, because it
does not allow to define a notion of energy.

Under gauge transformations generated by \eqref{eq:Generators-1-1},
the fields $g_{ij}$ and $\pi^{ij}$ transform as
\begin{equation}
\delta g_{ij}=\frac{2\xi}{\sqrt{g}}\left(\pi_{ij}-\frac{1}{2}g_{ij}\pi\right)+\xi_{i\mid j}+\xi_{j\mid i}\,,\label{eq:transfgelectric}
\end{equation}

\begin{eqnarray}
\delta\pi^{ij} & = & \frac{\xi}{2\sqrt{g}}g^{ij}\left(\pi^{kl}\pi_{kl}-\frac{1}{2}\pi^{2}\right)-\frac{2\xi}{\sqrt{g}}\left(\pi_{\;l}^{i}\pi^{lj}-\frac{1}{2}\pi\pi^{ij}\right)\nonumber \\
 &  & +\left(\xi^{k}\pi^{ij}\right)_{\mid k}-\xi_{\mid k}^{i}\pi^{kj}-\xi_{\mid k}^{j}\pi^{ki}\,.\label{eq:transfpielectric}
\end{eqnarray}

\subsection{Asymptotic behavior of the fields}

\subsubsection{Fall-off of the canonical variables}

To describe the asymptotic behavior of the canonical variables, we
will consider deviations with respect to the same Carrollian background
configuration used in the magnetic theory in eq. \eqref{eq:background}.
Since the momenta vanish for this configuration, the constraints of
the electric theory in eq. \eqref{eq:constraints_electric} are automatically
fulfilled. The fall-off of the fields must then guarantee that the
symplectic term is finite and that the charges are finite and integrable
when appropriate parity conditions are implemented. This is achieved
with exactly the same fall-off used in the case of the magnetic contraction
in eqs. \eqref{eq:grr}-\eqref{eq:pAB}, i.e.,

\begin{align}
g_{rr} & =1+\frac{f_{rr}}{r}+\frac{f_{rr}^{\left(-2\right)}}{r^{2}}+\mathcal{O}\left(r^{-3}\right)\,,\label{eq:grr-1}\\
g_{rA} & =\frac{f_{rA}^{\left(-1\right)}}{r}+\mathcal{O}\left(r^{-2}\right)\,,\label{eq:grA-1}\\
g_{AB} & =r^{2}\gamma_{AB}+r\,f_{AB}+f_{AB}^{\left(0\right)}+\mathcal{O}\left(r^{-1}\right)\,,\label{eq:gAB-1}
\end{align}
\begin{align}
\pi^{rr} & =p^{rr}+\mathcal{O}\left(r^{-1}\right)\,,\label{eq:prr-1}\\
\pi^{rA} & =\frac{p^{rA}}{r}+\frac{p_{\left(-2\right)}^{rA}}{r^{2}}+\mathcal{O}\left(r^{-3}\right)\,,\label{eq:prA-1}\\
\pi^{AB} & =\frac{p^{AB}}{r^{2}}+\mathcal{O}\left(r^{-3}\right)\,.\label{eq:pAB-1}
\end{align}
This asymptotic form is preserved under the transformations \eqref{eq:transfgelectric}-\eqref{eq:transfpielectric}
with gauge parameters of the form
\begin{align}
\xi & =r\,b+f+\mathcal{O}\left(r^{-1}\right)\,,\label{eq:xi-1}\\
\xi^{r} & =W+\mathcal{O}\left(r^{-1}\right)\,,\label{eq:xir-1}\\
\xi^{A} & =Y^{A}+\frac{1}{r}\left(\frac{2b}{\sqrt{\gamma}}p^{rA}+D^{A}W\right)+\mathcal{O}\left(r^{-2}\right)\,,\label{eq:xiA-1}
\end{align}
with
\begin{equation}
Y^{A}=\frac{\epsilon^{AB}}{\sqrt{\gamma}}\partial_{B}\left(\vec{\omega}\cdot\hat{r}\right)\,.\label{bY-1}
\end{equation}
Here, the functions $b=b\left(\theta,\phi\right)$, $f=f\left(\theta,\text{\ensuremath{\phi}}\right)$
and $W=W\left(\theta,\phi\right)$ are general arbitrary scalar functions
on the 2-sphere.

\subsubsection{Variation of the charge}

The conserved charges are obtained by replacing the asymptotic form
of the fields in eq. \eqref{eq:DeltaQE}. There is a naively divergent
term of the form

\[
Q_{E}^{\text{div}}=\int d^{2}x\;2\,Y_{A}\,p^{rA}\,,
\]
that vanishes provided the order $\mathcal{O}\left(1\right)$ of the
constraint $\mathcal{H}_{A}^{M}$, given by 
\[
p^{rA}+D_{B}p^{AB}=0\,,
\]
is implemented. Therefore, the charge is finite and integrable in
the functional sense. It takes the form:
\begin{equation}
Q_{E}=\oint d^{2}x\left[2Y^{A}\left(p^{rr}f_{rA}+p_{\;\,A}^{r\left(-2\right)}+p^{rB}f_{AB}\right)+\frac{2b}{\sqrt{\gamma}}\left(p^{rA}p_{\;\,A}^{r}\right)+2W\left(p^{rr}-2D^{A}p_{rA}\right)\right]\,.\label{eq:QE}
\end{equation}
Note that the parameter $b$ enters in this expression through the
order $\mathcal{O}\left(r^{-1}\right)$ of $\xi^{A}$, that is determined
by the gauge condition that requires that the order $\mathcal{O}\left(1\right)$
of $g_{rA}$ vanishes. As shown later, once parity conditions are
imposed, all the terms containing $b$ will no contribute to the charge,
and consequently they define pure gauge transformations (proper gauge
transformations).

\subsubsection{Transformation laws}

The transformation laws of the fields that are relevant for the parity
conditions are given by

\begin{equation}
\delta f_{rr}=\mathcal{L}_{Y}f_{rr}+\frac{b}{\sqrt{\gamma}}\left(p^{rr}-\tilde{p}\right)\,,\label{eq:DeltafrrEl}
\end{equation}
\begin{align}
\delta f_{AB} & =\mathcal{L}_{Y}f_{AB}+\frac{2b}{\sqrt{\gamma}}\left(p_{AB}-\frac{1}{2}\gamma_{AB}\left(\tilde{p}+p^{rr}\right)\right)+2\left(D_{A}D_{B}+\gamma_{AB}\right)W\nonumber \\
 & +\frac{2}{\sqrt{\gamma}}\left(\left(D_{A}b\right)\gamma_{BC}p^{rC}+\left(D_{B}b\right)\gamma_{AC}p^{rC}\right)+\frac{2b}{\sqrt{\gamma}}\left(\gamma_{BC}D_{A}p^{rC}+\gamma_{AC}D_{B}p^{rC}\right)\,,\label{eq:DeltafABEl}
\end{align}
\begin{equation}
\delta p^{rr}=\mathcal{L}_{Y}p^{rr}\,,\label{eq:DeltaprrEl}
\end{equation}
\begin{equation}
\delta p^{rA}=\mathcal{L}_{Y}p^{rA}\,,\label{eq:DeltaprAEl}
\end{equation}
\begin{equation}
\delta p^{AB}=\mathcal{L}_{Y}p^{AB}\,.\label{eq:DeltapABEL}
\end{equation}
Note that the momenta only transform under spatial rotations parametrized
by the vector field $Y^{A}$.

\subsubsection{Symplectic term}

The symplectic term takes exactly the same form as in the magnetic
case. In particular, the logarithmic divergence takes the form
\begin{equation}
\int dtd^{3}x\,\pi^{ij}\dot{g}_{ij}\underset{r\rightarrow\infty}{=}\log\left(r\right)\int dt\oint d^{2}x\left(p^{rr}\dot{f}_{rr}+p^{AB}\dot{f}_{AB}\right)+\mathcal{O}\left(r^{-1}\right)\,.\label{eq:Sympmag-2}
\end{equation}

\subsection{Asymptotic symmetries with Regge-Teitelboim parity conditions}

\subsubsection{Parity conditions}

As in the magnetic case, the Regge-Teitelboim parity conditions defined
in eqs. \eqref{eq:parityRTM1} and \eqref{eq:parityRTM2}, render
the symplectic term finite by eliminating the logarithmic divergence
in \eqref{eq:Sympmag-2}. The consistency of the analysis then requires
that the parity conditions are preserved under the transformation
laws of the fields \eqref{eq:DeltafrrEl}-\eqref{eq:DeltapABEL}.
In particular, from the transformation of $f_{rr}$ in eq. \eqref{eq:DeltafrrEl}
one finds that $b\left(\theta,\phi\right)$ must be odd. The transformation
law of $f_{AB}$ in \eqref{eq:DeltafABEl} implies that $W\left(\theta,\phi\right)$
is even under the antipodal map, with the exception of the mode that
is annihilated by the operator $\left(D_{A}D_{B}+\gamma_{AB}\right)$,
that corresponds to the mode with $\ell=1$ in the spherical harmonics
expansion. Therefore, we have the following restrictions on the parameters:
\[
b=b_{\text{odd}}\left(\theta,\phi\right)\,,\qquad\qquad W\left(\theta,\phi\right)=\vec{\alpha}\cdot\hat{r}+W_{\text{even}}\left(\theta,\phi\right)\,.
\]

\subsubsection{Charges and asymptotic symmetry algebra}

When Regge-Teitelboim parity conditions are imposed, the charge \eqref{eq:QE}
takes the form

\begin{equation}
Q_{E}=\oint d^{2}x\left[2\sqrt{\gamma}\,\vec{\omega}\cdot\hat{r}\epsilon_{AB}D^{A}p_{\left(-2\right)}^{rB}+2\,\vec{\alpha}\cdot\hat{r}\left(p^{rr}-2\tilde{p}\right)\right]\,.\label{eq:QERT}
\end{equation}
The only parameters that appear in $Q_{E}$ are $\vec{\omega}$ and
$\vec{\alpha}$, associated with spatial rotations and spatial translations,
respectively. The charge does not depend on the parameters associated
with normal deformations, in particular on the parameter describing
the time evolution. The origin of this property can be traced back
to the absence of spatial derivatives in the constraint $\mathcal{H}^{E}$
in eq. \eqref{eq:constraints_electric}, which in turn implies that
the boundary term associated with the Hamiltonian does not depend
on the lapse function, as it can be seen from eq. \eqref{eq:DeltaQE}.

The asymptotic symmetry algebra can then be obtained using eqs. \eqref{eq:complaw1}
and \eqref{eq:complaw2} with the following asymptotic form of the
parameters
\begin{align}
\xi & =\mathcal{O}\left(r^{-1}\right)\,,\label{eq:xi-1-1}\\
\xi^{r} & =\vec{\alpha}\cdot\hat{r}+\mathcal{O}\left(r^{-1}\right)\,,\label{eq:xir-1-1}\\
\xi^{A} & =\frac{\epsilon^{AB}}{\sqrt{\gamma}}\partial_{B}\left(\vec{\omega}\cdot\hat{r}\right)+\frac{1}{r}D^{A}\left(\vec{\alpha}\cdot\hat{r}\right)+\mathcal{O}\left(r^{-2}\right)\,,\label{eq:xiA-1-1}
\end{align}
where only the parameters associated with improper (or large) gauge
transformations were included. The algebra then closes according to
\[
\omega_{3}^{K}=-\epsilon_{IJK}\omega_{1}^{I}\omega_{2}^{J}\;,\qquad\alpha_{3}^{K}=-\epsilon_{IJK}\left(\alpha_{1}^{I}\omega_{2}^{J}-\alpha_{2}^{I}\omega_{1}^{J}\right)\,.
\]
Therefore, the asymptotic symmetry algebra in the electric Carrollian
theory with Regge-Teitelboim parity conditions is given by the semi-direct
sum of spatial rotations and spatial translations.

The canonical generators then read
\[
J_{I}=2\oint d^{2}x\,\sqrt{\gamma}\,\hat{r}_{I}\epsilon_{AB}D^{A}p_{\left(-2\right)}^{rB}\;,\qquad P_{I}=2\oint d^{2}x\,\hat{r}_{I}\left(p^{rr}-2\tilde{p}\right)\,.
\]
Thus, the charge \eqref{eq:QERT} takes the form 
\[
Q_{E}=\vec{\omega}\cdot\vec{J}+\vec{\alpha}\cdot\vec{P}\,.
\]
The generators then obey the following algebra in terms of the Poisson
brackets

\[
\left\{ J_{I},J_{J}\right\} =-\epsilon_{IJK}J_{K}\;,\qquad\left\{ P_{I},J_{J}\right\} =-\epsilon_{IJK}P_{K}\;,\qquad\left\{ P_{I},P_{J}\right\} =0\,.
\]
This is a subalgebra of the Carroll algebra that excludes the generators
associated with energy and boosts.

\subsection{Asymptotic symmetries with Henneaux-Troessaert parity conditions}\label{sec:HT-electric}

\subsubsection{Parity conditions}

The parity conditions introduced by Henneaux and Troessaert \cite{Henneaux:2018cst}
can also be implemented in the case of Electric Carrollian gravity.
As it was discussed in section \ref{subsec:Asymptotic-symmetries-with},
it is necessary to introduce the new set of variables $\bar{\lambda},\,\bar{k}_{AB},\,\bar{k},\,\bar{p}$
and $k^{\left(2\right)}$ defined in eq. \eqref{eq:newvar}. In terms
of these variables, the logarithmic divergence of the symplectic term
in eq. \eqref{eq:Sympmag-2} takes exactly the same form as the one
obtained in the case of the magnetic contraction in \eqref{eq:Sympmag-1}.
Then, once the parity conditions in \eqref{eq:parityHTM1} and \eqref{eq:parityHTM2}
are imposed, the logarithmic divergence is removed and the symplectic
term becomes finite.

In terms of the new variables, the transformation law of the fields
in eqs. \eqref{eq:DeltafrrEl}-\eqref{eq:DeltapABEL} now read

\begin{equation}
\delta\bar{\lambda}=\mathcal{L}_{Y}\bar{\lambda}+\frac{b}{4\sqrt{\gamma}}\bar{p}\,,\label{eq:deltalambdael}
\end{equation}
\begin{align}
\delta\bar{k}_{AB} & =\mathcal{L}_{Y}\bar{k}_{AB}+\frac{b}{\sqrt{\gamma}}\left(p_{AB}-\gamma_{AB}\bar{p}\right)+\left(D_{A}D_{B}+\gamma_{AB}\right)W\nonumber \\
 & +\frac{1}{\sqrt{\gamma}}\left(D_{A}\left(b\,p_{\;\,B}^{r}\right)+D_{B}\left(b\,p_{\;\,A}^{r}\right)\right)\,,\label{eq:deltakabel}
\end{align}
\begin{equation}
\delta\bar{p}=\mathcal{L}_{Y}\bar{p}\,,\label{eq:deltapel}
\end{equation}

\begin{equation}
\delta p^{rA}=\mathcal{L}_{Y}p^{rA}\,,\label{eq:deltaprAel}
\end{equation}

\begin{equation}
\delta p^{AB}=\mathcal{L}_{Y}p^{AB}\,.\label{eq:DeltapABelHT}
\end{equation}
The parity conditions of Henneaux and Troessaert must then be preserved
under the above gauge transformations. In particular, from the transformation
law of $\bar{\lambda}$ in eq. \eqref{eq:deltalambdael} one finds
that $b\left(\theta,\phi\right)$ must be odd under the antipodal
map on the 2-sphere. The consistency in the transformation of $\bar{k}_{AB}$
requires that $W$ is also parity odd. Note that $f\left(\theta,\phi\right)$
does not appear in these transformations, and consequently it does
not have any particular parity. In sum, the parameters $b$ and $W$
obey the following parity conditions
\begin{align*}
b\left(\theta,\phi\right) & \qquad\left(\text{parity odd}\right)\,,\\
W\left(\theta,\phi\right) & \qquad\left(\text{parity odd}\right)\,.
\end{align*}

\subsubsection{Charges and asymptotic symmetry algebra}

With Henneaux-Troessart parity conditions, the charge \eqref{eq:QE}
simplifies to

\begin{equation}
Q_{E}=\oint d^{2}x\;\left[W\,\bar{p}+2\left(\vec{\omega}\cdot\hat{r}\right)\frac{\epsilon^{AB}}{\sqrt{\gamma}}D_{A}\left(p_{\left(-2\right)}^{rB}-2\bar{\lambda}\,p^{rB}\right)\right]\,.\label{eq:QElHT}
\end{equation}
It depends only on $\vec{\omega}$ and $W\left(\theta,\phi\right)$
that correspond to the parameters of spatial rotations and ``parity
odd supertranslations,'' respectively. As in the case of Regge-Teitelboim
parity conditions, the parameters $b\left(\theta,\phi\right)$ and
$f\left(\theta,\phi\right)$ do not appear in the charges, and therefore
they are pure gauge.

If we consider the asymptotic form of the gauge parameters, where
only those associated with non-trivial (improper) symmetries are included,
one has
\begin{align}
\xi & =\mathcal{O}\left(r^{-1}\right)\,,\label{eq:xi-1-2}\\
\xi^{r} & =W+\mathcal{O}\left(r^{-1}\right)\,,\label{eq:xir-1-2}\\
\xi^{A} & =\frac{\epsilon^{AB}}{\sqrt{\gamma}}\partial_{B}\left(\vec{\omega}\cdot\hat{r}\right)+\frac{1}{r}D^{A}W+\mathcal{O}\left(r^{-2}\right)\,.\label{eq:xiA-1-2}
\end{align}
The asymptotic symmetry algebra can then be obtained using eqs. \eqref{eq:complaw1}
and \eqref{eq:complaw2}. It is characterized by
\begin{align*}
\omega_{K}^{3} & =-\epsilon_{IJK}\omega_{1}^{I}\omega_{2}^{J}\,,\\
W_{3} & =Y_{1}^{A}D_{A}W_{2}-Y_{2}^{A}D_{A}W_{1}\,.
\end{align*}
This algebra corresponds to the semi-direct sum of spatial rotations
with ``parity odd supertranslations'' parametrized by $\vec{\omega}$
and $W\left(\theta,\phi\right)$, respectively. The corresponding
canonical generators associated with the above symmetry transformations
are then given by
\begin{equation}
J_{I}=2\oint d^{2}x\;\sqrt{\gamma}\epsilon_{AB}\hat{r}_{I}\,D^{A}\left(p_{\left(-2\right)}^{rB}-2\bar{\lambda}p^{rB}\right)\;,\qquad\mathcal{P}\left(\theta,\phi\right)=\frac{\bar{p}}{\sqrt{\gamma}}\,,\label{eq:GenMagHT2-1}
\end{equation}
and are defined in such a way that the charge \eqref{eq:QElHT} takes
the form
\[
Q_{E}=\vec{\omega}\cdot\vec{J}+\oint d^{2}x\;\sqrt{\gamma}\,W\left(\theta,\phi\right)\,\mathcal{P}\left(\theta,\phi\right)\,.
\]
They fulfill the following algebra in terms of Poisson brackets:

\[
\left\{ J_{I},J_{J}\right\} =-\epsilon_{IJK}J_{K},\qquad\left\{ \mathcal{P}\left(\theta,\phi\right),J_{I}\right\} =\hat{Y}_{I}^{A}\partial_{A}\mathcal{P}\left(\theta,\phi\right),\qquad\left\{ \mathcal{P}\left(\theta,\phi\right),\mathcal{P}\left(\theta',\phi'\right)\right\} =0\,,
\]
where $\hat{Y}_{I}^{A}:=\frac{\epsilon^{AB}}{\sqrt{\gamma}}\partial_{B}\hat{r}_{I}$.
Here $J_{I}$ denotes the generators of spatial rotations, while the
odd modes of $\mathcal{P}\left(\theta,\phi\right)$ correspond to
the ``parity odd supertranslations.'' In particular, the standard
spatial translations are given by the modes with $\ell=1$ in the
spherical harmonics expansion.

\subsection{Example: Electric Carrollian Schwarzschild-like solution}

In the electric case, one can also find a configuration that solves
the constraints \eqref{eq:constraints_electric} and the equations
of motion \eqref{eq:HamEl1}, \eqref{eq:Eqpi} with the same spatial
metric components that the Schwarzschild solution in General Relativity.
\[
g_{rr}=\frac{1}{\left(1-\frac{M}{8\pi r}\right)}\;,\qquad\qquad g_{rA}=0\;,\qquad\qquad g_{AB}=r^{2}\gamma_{AB}\,,
\]
\[
\pi^{rr}=0\;,\qquad\qquad\pi^{rA}=0\;,\qquad\qquad\pi^{AB}=0\,.
\]
From the dynamical equations \eqref{eq:HamEl1}, \eqref{eq:Eqpi},
for a spherically symmetric ansatz one finds
\[
N=N\left(r\right)\;,\qquad\qquad N^{A}=0\,.
\]
Note that in this case the lapse is not determined by the equations
of motion. This is in line with the fact that normal surface deformations
do not appear in the charges and that therefore are ``pure gauge.''
The spatial components of the metric then admit the following asymptotic
expansion:
\begin{align*}
g_{rr} & =1+\frac{M}{8\pi r}+\frac{M^{2}}{64\pi^{2}r^{2}}+\dots\,,\\
g_{AB} & =r^{2}\gamma_{AB}\,.
\end{align*}
This behavior fits within the fall-off in eqs. \eqref{eq:grr-1}-\eqref{eq:pAB-1},
and also obeys the Regge-Teitelboim parity conditions. However, in
contrast to the magnetic contraction, there are no charges associated
with this configuration because $f_{rr}$ does not appear in the charges,
and consequently, it can be consistently gauged away.

A similar situation occurs in the case of Henneaux-Troessaert parity
conditions, where after performing a radial shift $r\rightarrow r+M/\left(16\pi\right)$,
one obtains that $\bar{\lambda}=M/(16\pi)$. Again, $\bar{\lambda}$
does not appear in the charges, and consequently there are no charges
associated with this configuration.

\section{Concluding remarks\label{sec:5 Final-remarks}}

The asymptotic symmetries of the magnetic and electric Carrollian
contractions of General Relativity were analyzed in 3+1 space and
time dimensions. The electric gravitational theory was known from
long ago as a ``strong coupling limit'' \cite{Isham:1975ur} or
as a``zero signature limit'' \cite{Teitelboim:1978wv} of General
Relativity, while the magnetic one was recently introduced by Henneaux
and Salgado-Rebolledo in ref. \cite{Henneaux:2021yzg} in the context
of the study of Carrollian contractions of generic relativistic theories.
These Carrollian gravities are described in the Hamiltonian formalism,
and the Regge-Teitelboim method was used to determine the asymptotic
symmetry algebra, together with their corresponding canonical generators.\footnote{The electric gravitational theory was described in a manifestly covariant
form under local Carroll symmetries in ref. \cite{Henneaux:1979vn}
(see e.g. \cite{Henneaux:2021yzg} for a review).} In order to render the symplectic term finite, it was necessary to
implement appropriate parity conditions on the leading order of the
canonical variables, in complete analogy with the case of General
Relativity. Two possibilities were explored: Regge-Teitelboim parity
conditions \cite{Regge:1974zd} and Henneaux-Troessaert parity conditions
\cite{Henneaux:2018cst}.

In the magnetic theory, for Regge-Teitelboim parity conditions, the
asymptotic symmetry algebra corresponds to the finite-dimensional
Carroll algebra of Levy-Leblond \cite{levy1965nouvelle,Bacry:1968zf},
that can be obtained from an ``ultrarelativistic'' contraction $(c\rightarrow0)$
of the Poincar\'e algebra. On the other hand, when Henneaux-Troessaert
parity conditions are imposed, the asymptotic symmetry algebra is
given by a ``BMS-like'' extension of the Carroll algebra by an infinite
number of ``Carrollian supertranslations.'' This infinite-dimensional
algebra can be recovered from a contraction of the BMS$_{4}$ algebra
written in the non-standard basis used by Henneaux and Troessaert
in ref. \cite{Henneaux:2018cst}. It would be interesting to try to
make contact with a possible ultrarelativistic contraction of the
BMS$_{4}$ algebra in the standard basis used by Sachs \cite{Sachs:1962},
and determine if it can be obtained directly from an asymptotic analysis
of what would be the analog of null infinity in this Carrollian context.
A possibility could be the use of the so called ``asymptotically
null space-like hypersurfaces'' introduced in refs. \cite{Bunster:2018yjr,Bunster:2019mup}
in the study of asymptotic symmetries at null infinity in General
Relativity using the canonical formalism.

For the theory obtained from the electric contraction, the asymptotic
symmetry algebra is truncated to the semi-direct sum of spatial rotations
and spatial translations for Regge-Teitelboim parity conditions, and
to the semi-direct sum of spatial rotations and an infinite number
of parity odd supertranslations in the case of Henneaux-Troessaert
parity conditions. In both cases, there are no canonical generators
associated with normal displacements to the hypersurfaces defined
at constant times in the asymptotic region, that include the time
evolution at the boundary. Therefore, these transformations are pure
gauge, and consequently there is no notion of energy in this theory.
This very special property can be traced back to the absence of spatial
derivatives in the Hamiltonian constraint that does not lead to boundary
terms associated with it.

These results show a discontinuity between the case of Einstein gravity
and its strict electric Carrollian limit. In General Relativity with
Regge-Teitelboim parity conditions, the asymptotic symmetry algebra
is the ten-dimensional Poincar\'e algebra \cite{Regge:1974zd}, even
if it is treated perturbatively in the signature parameter \cite{Teitelboim:1983fi}.
On the other hand, in the strict electric Carrollian limit, the asymptotic
symmetries with Regge-Teitelboim parity conditions are described by
the six-dimensional algebra spanned by spatial rotations and spatial
translations. It possesses less generators than the original one found
before the limit. This is particularly interesting because the electric
limit was precisely introduced as the zero order in a perturbative
expansion in terms of the ``signature parameter.'' This suggests
that this jump in the symmetry must be carefully taken into account
in this perturbative approach when boundaries are present. A similar
situation occurs for Henneaux-Troessaert parity conditions but in
terms of the corresponding extended symmetry algebras. On the contrary,
the ultrarelativistic limit in the magnetic contraction is perfectly
smooth for both parity conditions, at least from the point of view
of the asymptotic symmetries \footnote{In General Relativity there is an additional set of boundary
 conditions that leads to the BMS algebra at spatial infinity \cite{Henneaux:2018hdj,Henneaux:2019yax}.
  The implementation of the analog of these boundary conditions in the Carrollian case was not analyzed in this article. 
  However, Prof. Marc Henneaux pointed out to me that the conclusions are the same as those obtained using Henneaux-Troessaert
   parity conditions in sections \ref{subsec:Asymptotic-symmetries-with} and \ref{sec:HT-electric}. I thank Prof. Henneaux for this point.}.

Different gravitational Carrollian theories were obtained in refs.
\cite{Hartong:2015xda,Bergshoeff:2017btm} by considering the gauging
of the Carroll algebra. Despite the fact that the relation with the
Carrollian gravities obtained from electric and magnetic contractions
of General Relativity is not known, in ref. \cite{Henneaux:2021yzg}
it was conjectured that the action constructed in \cite{Hartong:2015xda}
could be related with the electric contraction, while the action introduced
in \cite{Bergshoeff:2017btm} with the magnetic one. It would be an
interesting problem to explore the asymptotic symmetries of the Carrollian
gravities of \cite{Hartong:2015xda,Bergshoeff:2017btm} because, in
principle, they could help to understand the possible relations between
the different Carrollian gravitational theories.

The results reported in this article could also be extended to include
a negative cosmological constant. As it was pointed out in ref. \cite{Henneaux:2021yzg},
it is straightforward to modify the action principle of the magnetic
and electric contractions to incorporate it. However, in contrast
to the case with a vanishing cosmological constant, both theories
do not share the same Carrollian background configuration. For example,
the electric contraction cannot have a background solution with vanishing
momenta because this configuration is forbidden by the Hamiltonian
constraint. This fact generically leads to striking differences in
comparison with the case without a cosmological constant \cite{Perez2021}.

The generalization to higher dimensions, as well as the coupling to
electric/magnetic Carrollian matter fields could also be explored.

\acknowledgments{I wish to thank Claudio Bunster, Joaquim Gomis, Marc Henneaux, Stefan Prohazka, Ricardo Troncoso and Jakob Salzer for some useful comments.  I also thank the hospitality of the Erwin Schrodinger Institute (ESI) in Vienna during the program ``Geometry for Higher Spin Gravity: Conformal Structures, PDEs, and Q-manifolds.'' I am also grateful with Daniel Grumiller for his kind hospitality at the TU-Wien in Vienna, and with  Stefan Theisen for his warm hospitality at the Albert Einstein Institute in Golm.  This research has been partially supported by Fondecyt grants N$\textsuperscript{\underline{o}}$  1171162, 1181031, 1181496, 1211226.}

\appendix

\section{Appendix A: Analysis in Cartesian coordinates with Regge-Teitelboim
parity conditions \label{sec:Appendix A}}

The results obtained using Regge-Teitelboim parity conditions are
reproduced in Cartesian coordinates. This approach resembles the original
one introduced by Regge and Teitelboim in General Relativity \cite{Regge:1974zd}.

Let us consider the same fall-off for the canonical variables as the
one used originally by Regge and Teitelboim 
\begin{equation}
g_{ij}=\delta_{ij}+\frac{h_{ij}^{\left(1\right)}\left(\hat{r}\right)}{r}+\frac{h_{ij}^{\left(2\right)}}{r^{2}}+\dots\,,\label{eq:Fall-offgCartesian}
\end{equation}
\begin{equation}
\pi^{ij}=\frac{p_{\left(2\right)}^{ij}\left(\hat{r}\right)}{r^{2}}+\frac{p_{\left(3\right)}^{ij}}{r^{3}}+\dots\,.\label{eq:Fall-offpiCartesian}
\end{equation}
In this section the coordinates $x^{i}$, with $i=1,2,3$, denote
Cartesian coordinates.

In terms of these variables, the Regge-Teitelboim parity conditions
take the form
\begin{equation}
h_{ij}^{\left(1\right)}\left(\hat{r}\right)=h_{ij}^{\left(1\right)}\left(-\hat{r}\right)\;,\qquad\qquad p_{\left(2\right)}^{ij}\left(\hat{r}\right)=-p_{\left(2\right)}^{ij}\left(-\hat{r}\right)\,.\label{eq:Parity_cond:Cartesian}
\end{equation}
The logarithmic divergence of the symplectic is then given by
\[
\int dtd^{3}x\,\pi^{ij}\dot{g}_{ij}\underset{r\rightarrow\infty}{=}\log\left(r\right)\oint d\theta d\phi\sin\theta\,p_{\left(2\right)}^{ij}\dot{h}_{ij}^{\left(1\right)}+\dots\,.
\]
By virtue of the parity conditions \eqref{eq:Parity_cond:Cartesian},
the divergence is removed and the symplectic term becomes finite.

\subsection{Magnetic Carrollian gravity}

The asymptotic Killing vectors associated with improper gauge transformations
can be written as
\[
\xi=\beta_{i}x^{i}+T+\dots\;,\qquad\qquad\xi_{i}=\epsilon_{ijk}x^{j}\omega^{k}+\alpha_{i}+\dots\,.
\]
In spherical coordinates these expressions coincide with the ones
in eqs. \eqref{eq:xi-2}-\eqref{eq:xiA-2}.

\emph{Energy:}

The energy is obtained for $\xi\underset{r\rightarrow\infty}{\longrightarrow}1$
and $\xi^{i}\underset{r\rightarrow\infty}{\longrightarrow}0$ in $\delta Q_{M}$
given by \eqref{eq:deltaQ}. Thus
\[
E=\oint d^{2}s_{k}\left(g_{ki\mid i}-g_{ii\mid k}\right)\,.
\]
 Note that the expression for the energy in the magnetic Carrollian
theory coincides with the formula for the ADM mass in General Relativity
\cite{Arnowitt:1962hi}.

\emph{Momentum:}

The momentum is obtained for $\xi\underset{r\rightarrow\infty}{\longrightarrow}0$
and $\xi^{i}\underset{r\rightarrow\infty}{\longrightarrow}\alpha^{i}$
\[
Q_{M}\left[\vec{\alpha}\right]=2\alpha_{k}\oint d^{2}s_{i}\pi^{ik}=\alpha_{k}P^{k}\,.
\]

\emph{Angular momentum}:

The angular momentum is obtained for $\xi\underset{r\rightarrow\infty}{\longrightarrow}0$
and $\xi_{i}\underset{r\rightarrow\infty}{\longrightarrow}\epsilon_{ijk}x^{j}\omega^{k}$

\[
Q_{M}\left[\vec{\omega}\right]=2\omega^{i}\oint d^{2}s_{l}\epsilon_{ijk}\pi^{lj}x^{k}=\omega^{i}J_{i}\,.
\]

\emph{Boosts:}

Boost generators are obtained when $\xi\underset{r\rightarrow\infty}{\longrightarrow}\beta_{k}x^{k}$
and $\xi^{i}\underset{r\rightarrow\infty}{\longrightarrow}0$
\[
Q_{M}\left[\vec{\beta}\right]=\beta_{k}\oint d^{2}s_{l}\left[x^{k}\left(g_{il\mid i}-g_{ii\mid l}\right)-g_{kl}+\delta_{kl}g_{ii}\right]=\beta^{k}K_{k}\,.
\]

\subsection{Electric Carrollian gravity}

For the electric contraction, the fall-off of the canonical variables
takes exactly the same form as in the magnetic theory, given by eqs.
\eqref{eq:Fall-offgCartesian} and \eqref{eq:Fall-offpiCartesian}.

The asymptotic form of the gauge parameters that preserve the above
asymptotic conditions is then given by
\begin{align*}
\xi & =\mathcal{O}\left(r^{-1}\right)\,,\\
\xi_{i} & =\epsilon_{ijk}x^{j}\omega^{k}+\alpha_{i}+\dots\,.
\end{align*}
Here, only the terms associated with improper gauge transformations
were included.

\emph{Momentum:}

The momentum is obtained when $\xi\underset{r\rightarrow\infty}{\longrightarrow}0$
and $\xi^{i}\underset{r\rightarrow\infty}{\longrightarrow}\alpha^{i}$
\[
Q_{E}\left[\vec{\alpha}\right]=2\alpha_{k}\oint d^{2}s_{i}\pi^{ik}=\alpha_{k}P^{k}\,.
\]

\emph{Angular momentum}:

The angular momentum is obtained when $\xi\underset{r\rightarrow\infty}{\longrightarrow}0$
and $\xi_{i}\underset{r\rightarrow\infty}{\longrightarrow}\epsilon_{ijk}x^{j}\omega^{k}$
\[
Q_{E}\left[\vec{\omega}\right]=2\omega^{k}\oint d^{2}s_{i}\epsilon_{ljk}x^{j}\pi^{il}=\omega^{k}J_{k}\,.
\]

\bibliographystyle{../JHEP}
\bibliography{../review}

\end{document}